\documentclass[aps,prl,twocolumn,showpacs,superscriptaddress,groupedaddress]{revtex4}  
\usepackage{graphicx}  
\usepackage{dcolumn}   
\usepackage{natbib}
\usepackage{bm}        
\usepackage{amssymb}   
\newcommand{\bigO}{\ensuremath{\mathcal{O}}}
\begin{document}



\title{Transverse pulse shaping and optimization of a tapered hard X-ray free electron laser}
%
\affiliation{University of California, Los Angeles, California 90095, USA}
\affiliation{Stanford Linear Accelerator Center, Menlo Park, California 94025, USA}
\author{C.~Emma} \affiliation{University of California, Los Angeles, California 90095, USA}
\author{C.~Pellegrini} \affiliation{University of California, Los Angeles, California 90095, USA}\affiliation{Stanford Linear Accelerator Center, Menlo Park, California 94025, USA}
\author{J.~Wu} \affiliation{Stanford Linear Accelerator Center, Menlo Park, California 94025, USA}

%
%
%
\vskip 0.25cm
\date{\today}

\begin{abstract}
Multidimensional optimization schemes for TW hard X-Ray free electron lasers are applied to the cases of transversely uniform and parabolic electron beam distributions and compared to examples of transversely Gaussian beams. The optimizations are performed  for a $200$m undulator and a resonant wavelength of $\lambda_r=1.5\AA $ using the fully 3-dimensional FEL particle code GENESIS. Time dependent simulations showed that the maximum radiation power is larger for flatter transverse distributions due to enhanced optical guiding in the tapered section of the undulator. For a transversely Gaussian beam the maximum output power was found to be $\text{P}_{max}$=$1.56$ TW compared to $2.26$ TW for the parabolic case and $2.63$ TW for the uniform case. Spectral data also showed a 30-70$\%$ reduction in energy deposited in the sidebands for the uniform and parabolic beams compared with a Gaussian. An analysis of the maximum power as a function of detuning from resonance shows that redshifting the central wavelength by $\Delta \lambda / \lambda <\rho$ increases the power for all three transverse electron distributions. 
\end{abstract}

\maketitle


X-ray free-electron lasers (X-FELs) have given us for the last few years the capability of exploring the structures and dynamical processes of atomic and molecular systems with a simultaneous resolution in space and time of 1 $\AA$ and 1 fs.  LCLS, the most powerful existing X-FEL, delivers diffraction limited X-ray pulses in the energy range of 0.25 to 10 keV, with a pulse duration of a few to hundred femtoseconds, peak power at saturation of 20-30 GW and a line-width on the order of $10^{-3}$. SACLA, in Japan, gives similar characteristics \citep{proceedings}.

Many new scientific results have already been obtained using the present peak power level of X-FELs , in particular in bio-imaging and nonlinear science. This areas of research will benefit from a larger number of photons/pulse, a factor of ten or more within 10-20 fs, corresponding to a peak  power of 1 TW or more. 

The peak power can be increased by tapering the FEL undulator magnetic field to match the electron energy loss \citep{1071285}. Following this idea there has been recently much work to optimize the tapered section of a self-seeded X-FEL\citep{PhysRevSTAB.Y.Jiao}. The analytic models developed in these studies have considered electron beams with Gaussian transverse density profile. In this paper we investigate the effects of manipulating the transverse electron beam density profile by extending previous optimization methods to include transvesely parabolic, and the tranversely uniform electron distributions. The results are compared with those for a transversely Gaussian beam in both single frequency and time dependent GENESIS simulations of a 200-m, self seeded hard X-ray tapered FEL with LCLS-II-like parameters. 

As has been pointed out in recent work \citep{PhysRevSTAB.Y.Jiao}, 3 dimensional effects such as diffraction and refraction of the X-rays are key in determining the performance of self-seeded tapered FELs. Thus we examine first some of the central features of previously developed extensions to the one dimensional theory of Kroll, Morton and Rosenbluth \citep{1071285}, which describe the energy gain of the radiation field in a tapered FEL taking into account the importance of the transverse electron and radiation distributions. Starting from conservation of energy and applying the same assumptions as \citep{PhysRevSTAB.Y.Jiao} we can write the radiation power as a function of the longitudinal position in the undulator:

\begin{equation}
P(z)=\frac{\pi r_s(z)^2a_{s0}(z)^2}{4Z_0}\left(\frac{k_s m_e c^2}{e}\right)^2
\end{equation}

where $a_{s0}(z)=|e|A_s(z)/\sqrt{2} mc^2$ is the on-axis normalized vector potential of the radiation field for a linearly polarised undulator, $r_s(z)$ is the radiation beam size, $k_s$ is the radiation wavenumber and $Z_0$ is the impedance of free space. In order to maximize this output power we must therefore optimize the growth of the radiation field as a function of the longitudinal distance in the undulator. It is well known that this can be achieved through an adiabatic decrease in the resonant energy $\gamma_r mc^2$ of the electron beam \citep{1071285} where the resonant energy is given by:

\begin{equation}
\gamma_r^2(z)=\frac{k_w}{2k_s}\left(1+a_w(z)^2\right)
\end{equation}

where $k_w=2\pi/\lambda_w$ is the undulator wavenumber and $a_w(z)=|e|B_w(z)/\sqrt{2} k_wmc^2$ is the normalized vector potential of the undulator field which in the tapered case is a function of $z$. A decrease in  $\gamma_r $ can be obtained by varying both the undulator period and the undulator parameter, however in this study we will examine only the more convenient constant $k_w$ case. We consider applying the following taper profile post saturation \citep{PhysRevSTAB.Y.Jiao}:

\begin{figure}
\includegraphics[scale=0.4]{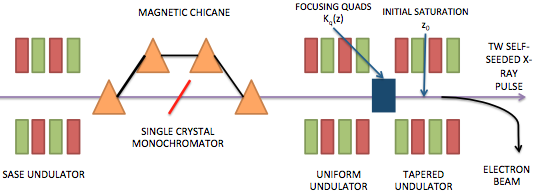}
\caption{Schematic representation of a tapered X-ray FEL using a self-seeding monochromator and an optimised tapered section}
\label{selfseeding}
\end{figure}

\begin{equation}
a_w(z)=a_w(z_0)\times[1-c\times(z-z_0)^d]
\end{equation}

where $z_0$ is the location of initial saturation (see figure \ref{selfseeding}) and $c$ and $d$ are constants to be obtained through simulations that maximize the output radiation power. After the initial saturation and exponential gain regime, the FEL process is dominated by the effects of refractive guiding of the radiation by the electron beam. This can be described quantitatively by considering the complex refractive index of the beam \citep{PhysRevSTAB.Y.Jiao}:

\begin{equation}
n=1+\frac{\omega_{p0}^2}{\omega_s^2}\frac{r_{b0}^2}{r_b^2}\frac{a_w}{2|a_s|}[JJ]\left\langle\frac{e^{-i\Psi}}{\gamma}\right\rangle
\label{rindex}
\end{equation}

where $\omega_p$ is the electron plasma frequency, $\omega_s$ is the radiation frequency and $r_b$ is the electron beam radius. Quantities with subscript 0 refer to initial parameters and the symbol $[JJ]=J_0(x)-J_1(x)$ for a planar undulator and $[JJ]=1$ for a helical undulator, where $x=a_w^2/2(1+a_w^2)$. The average term in square brackets is over the beam electrons where $\Psi$ is the individual electron phase relative to the ponderomotive potential. 

\begin{figure}
\includegraphics[scale=0.35]{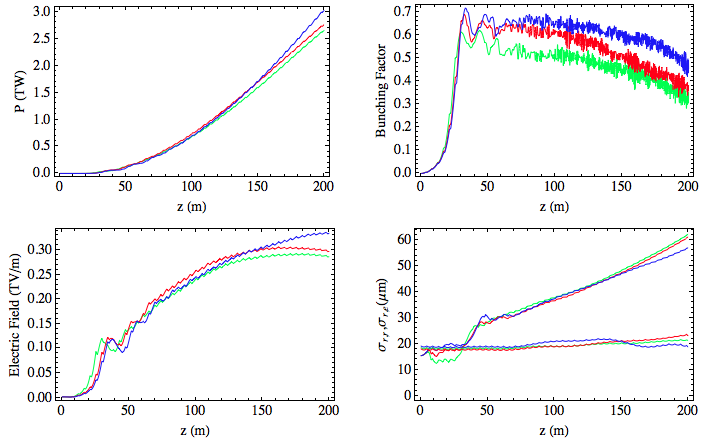}
\caption{Comparison of time independent results for Gaussian (green), parabolic (red) and uniform (blue) transverse beam distributions at $\lambda_r=1.5\AA$}
\label{tindpfigures}
\end{figure}

In the resonant particle approximation we assume that only trapped electrons contribute to the microbunching term $\left\langle \text{exp} (-i\Psi)\right\rangle$ and the radiation intensity increase. It is thus important to maintain a sufficiently large trapping fraction throughout the tapered region as this will boost the coherent interaction between the radiation and the electrons by increasing the refractive index of the electron beam. The increase in refractive index then provides optical guiding for the radiation field and limits the detrimental effects of diffraction\citep{PhysRevA.24.1436}. It is useful at this point to consider the function $F_t(z)$ which determines the fraction of electrons trapped in the FEL bucket along the tapered undulator:

\begin{equation}
F_t(z)=\frac{1}{N_e}\int_0^{r_{max}}F_t(r,z)f_0(r)2\pi r dr
\label{trappingfraction}
\end{equation}

where $f_0(r)$ is the transverse beam distribution and the \textit{local} trapping fraction $F_t(r,z)$ which is determined by the maximum and minimum phases $\Psi (r,z)$ for which particles follow stable trajectories in phase space (see \citep{PhysRevSTAB.Y.Jiao}\citep{1071285}).  It has been shown in \citep{PhysRevSTAB.Y.Jiao} that the microbunching term $\left\langle \text{exp}(-i\Psi)\right\rangle$ in eq \ref{rindex} can be calculated by averaging the product $F_t(r)$exp$[-i \Psi_r(r)]$ over the radial coordinate $r$. Examining equation \ref{trappingfraction} shows that by manipulating the initial transverse electron beam distribution $f_0(r)$ it is possible to maximize the trapping fraction throughout the tapered undulator. For the case of a transverse Gaussian distribution, considered thus far, the electrons in the tail of the beam experience a weaker electric field from those closer to the axis, and thus can become de-trapped from the FEL bucket, reducing the microbunching and causing an early saturation of the power gain inside the tapered undulator. If however the transverse electron distribution is tailored such that it is flatter with a narrower tail (such as the transversely parabolic or uniform cases) then it is possible to trap more electrons in the bucket, increasing the bunching factor throughout the undulator and thereby extracting more power. By inspecting equation \ref{rindex} and noting that the contribution from the Bessel functions $[JJ]$ is always less than 1 for planar undulators, we can also infer that the output power will increase in a helical undulator by a factor $\bigO{(1/[JJ])}$. 

\begin{figure}
\includegraphics[scale=0.25]{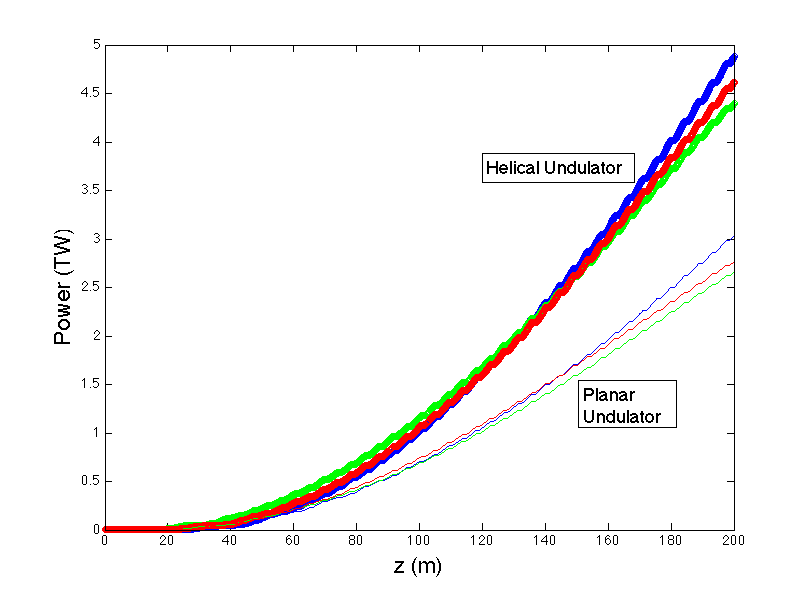}
\caption{Comparison of time independent results for Gaussian (green), parabolic (red) and uniform (blue) beams in planar and helical undulators with optimized taper profiles}
\label{helical}
\end{figure}
\begin{table}
\caption{\label{tab:table1}GENESIS Simulation Parameters: Single Frequency}
\begin{ruledtabular}
\begin{tabular}{lc}
Parameter Name&Parameter Value\\
\hline\hline
Beam Energy $E_0$ & 13.64 GeV \\
Beam Peak Current $I_{pk}$ & 4000 A \\
Normalized Emittances $\epsilon_{x,n}/\epsilon_{y,n}$ & 0.3/0.3 $\mu$ m rad \\
Peak radiation power input $P_{in}$ & 5 MW \\
Undulator Period $\lambda_w$ & 32 mm\\
Normalised Undulator Parameter$a_w$ & 2.3832\\
Radiation Wavelength $\lambda_r$ & 1.5 $\AA$\\
FEL parameter $\rho $ & 7.361 $\times 10^{-4}$ $ $\\
\end{tabular}
\end{ruledtabular}
\end{table}

\begin{figure*}[t]
\includegraphics[scale=0.45]{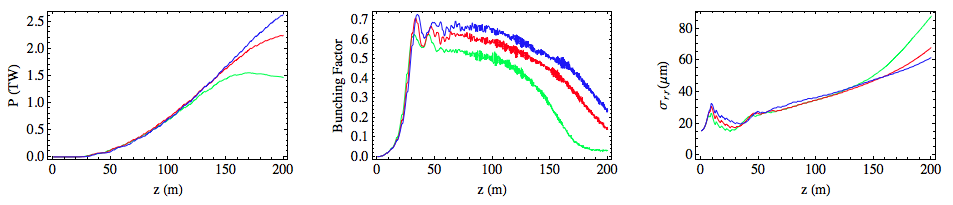}
\includegraphics[scale=0.45]{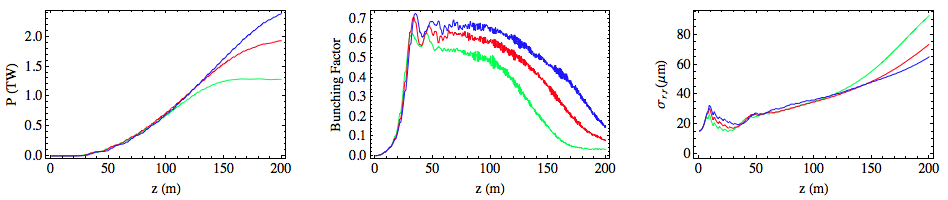}
\caption{Power, bunching factor and radiation size as a fucntion of longitudal distance for transversely Gaussian (green), parabolic (red) and uniform (blue) beams. The results are shown for a wavelength $\lambda_r = 1.5 \AA$ and bunch lengths of 6.4 fs (top) and 16 fs (bottom). The optimised taper profiles are found in time independent simulations.}\label{tdpfigure}
\end{figure*}


We first examine results from single frequency simulations performed with GENESIS which optimize the taper profile and transverse quadrupole focusing in a 200-m undulator with periodic break sections. The undulator parameters are similar to those of the LCLS-II upgrade project: 1-m long break sections, 3.4-m long undulator modules \citep{PhysRevSTAB.Y.Jiao}. As illustrated in figure \ref{tindpfigures} there is a noticeable increase in the bunching factor for the transversely parabolic and uniform distributions as compared to the Gaussian. This is indicative of a larger trapping fraction and consequently greater output power. The difference in maximum output power is however only marginally observed in single frequency simulations with the Gaussian beam achieving $\text{P}_{max}$=$2.65$ TW compared to $\text{P}_{max}$=$2.76$ TW for the parabolic case and $\text{P}_{max}$=$3.03$ TW for the uniform case. What is also apparent from figure \ref{tindpfigures} is that the radiation size is virtually unaffected by changing the transverse beam distribution, thus the effect of optical guiding is not observed to contribute significantly when multiple frequency effects are not included. 

We consider now the optimized taper profiles in the simulations shown in figure \ref{tindpfigures} and change the undulator geometry from planar to helical. The difference in output power is depicted in figure \ref{helical}, from which it is clear that the helical undulator shows an increase in the maximum power output of a factor 60-70 $\%$ for all three transverse electron distributions. Computing the Bessel factor $1/[JJ]=1.3$ reveals that this is due to the 30$\%$ increase in the refractive index of the beam. \newline

Using the optimial undulator parameters found via the time independent simulations, we performed time dependent simulations of the three transverse distributions for 6.4 fs and 16 fs bunch lengths. The results for a resonant wavelength $\lambda_r=1.5\AA $ are illustrated in figure \ref{tdpfigure}. From these it is clear that time dependent effects have a major impact on the output of an optimized tapered FEL and the 3-D effects of the transverse electron distribution are more significant when time dependence is included. Comparing figure \ref{tdpfigure} with the time independent results of figure \ref{tindpfigures} we notice that the uniform and parabolic distributions exhibit a  a steady growth in output power, and a slow decrease in the bunching factor throughout the tapered undulator. On the other hand the transversely Gaussian beam suffers a signficant reduction in the bunching and output power as well as an increased diffraction of the radiation. Furthermore, the transversely Gaussian beam shows an early saturation of the output power in the time dependent case, a result previously reported in \citep{PhysRevSTAB.Y.Jiao}. 

\begin{figure}
\includegraphics[scale=0.6]{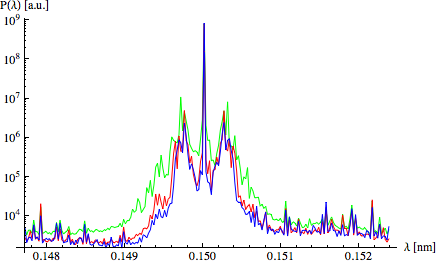}
\label{spectratdp}
\caption{Spectrum and sideband comparison for Gaussian (green) parabolic (red) and uniform (blue) transverse distributions at $\lambda_r=1.50078 \AA$.}
\end{figure}

Contributing to this disparity in maximum output power are the detrimental effects of the sideband instability. The flatter transverse profiles of the parabolic and uniform distributions mitigate more effectively the power losses and energy deposited in the sidebands than the transversely Gaussian beam. Evidence for this is shown in figure \ref{spectratdp} where observation of the spectrum at $\lambda_r=1.50078 \AA$ shows the Gaussian profile exhibits broader sidebands than the parabolic and uniform distributions. Examining the case for a 6.4 fs bunch length, we integrate the power deposited in the sidebands and notice a reduction in sideband energy of $~61 \%$ for the parabolic case and $~ 72 \%$ for the uniform beam as compared to the Gaussian. As is shown figure \ref{tdpfigure} this effect along with particle detrapping results in a sizeable difference in power output between the three transverse distributions with $\text{P}_{max}= 1.56$ TW for the Gaussian beam, $\text{P}_{max}= 2.26$ TW for the parabolic beam and $\text{P}_{max}= 2.63$ TW for the uniform beam. \newline

\begin{figure}
\includegraphics[scale=0.5]{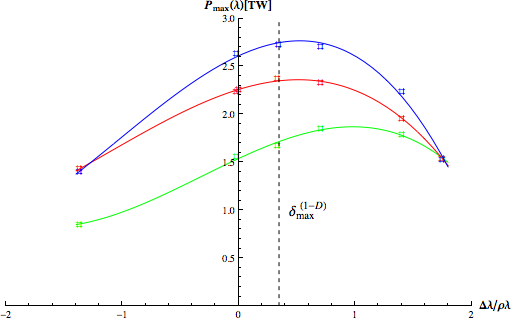}
\caption{Maximum radiation power for different values of detuning in transversely Gaussian (green), parabolic (red) and uniform (blue) beams}
\label{detuningcomparison}
\end{figure}

It is known from 3-D FEL theory of fixed parameter undulators that it is possible to maximize the output power by detuning the central wavelength of the beam with respect to the resonant wavelength \citep{PhysRevA.46.6662}. Here we examine this effect numerically in a tapered X-FEL by shifting the electron beam resonant wavelength to find the maximum of the small-signal high-gain curve. For the three transverse distributions with 6.4 fs bunch lengths, figure \ref{detuningcomparison} shows the gain curves obtained from multiple time dependent GENESIS simulations as a function of the normalized detuning $\delta=\Delta \lambda /\rho \lambda$. The maximal gain for the uniform and parabolic beams is found for a redshift of $\Delta \lambda / \lambda = 0.34 \rho$ while the Gaussian beam peaks at $\Delta \lambda / \lambda = 0.69 \rho$. The difference between the maximal detuning can be qualitatively understood by considering an extension to the 1-D FEL dispersion relation which includes the effect of refractive guiding. Neglecting energy spread effects we can write the dispersion relation as \citep{murphypellegrini}: 

\begin{equation}
\lambda^3-\left(1-\frac{\delta}{2\rho k_wZ_R}\lambda^2\right)+1=0
\end{equation}

where $\text{Im}(\lambda)$ is the growth rate of the field and $Z_R=\pi w_0^2/\lambda_r= 2\pi \sigma_{r,0}^2/\lambda_r$ is the Rayleigh range of the radiation for a beam of transverse rms size $\sigma_{r,0}$ at the waist. The 1-D estimate for the maximal growth rate is then achieved for a detuning $\delta_{max}^{(1-D)}=1/2k_w\rho Z_R$ which using the value of $\sigma_{r,0}= 15.5 \mu m$ employed in simulations yields $\delta_{max}^{(1-D)}=0.335\rho$ (see fig \ref{detuningcomparison}). While this value is in good agreement with the uniform and parabolic results it is off by approximately a factor of two for the Gaussian case. It is thus clear that in order to obtain a more complete understanding of this effect a fully 3-dimensional treatment is required since the Rayleigh range is in general a function of the transverse electron distribution. This analysis is currently being performed and results will be presented in future work. 
\newline

An extension to the multidimensional optimization of tapered free electron has been carried out to evaluate the effects of transverse electron beam shaping. The performance of previously considered transversely Gaussian beams was compared to transversely parabolic and uniform electron distributions. Optimizations were performed for a 200-m long undulator with break sections using the three dimensional particle code GENESIS in time dependent and time independent simulations. Time independent results show that the effect of changing the transverse beam distribution was mostly marginal, however when multifrequency effects are taken into account in time dependent simulations, the transverse distribution has an important impact on the FEL process affecting the trapping fraction and consequently the maximum output power. For a resonant wavelength of $\lambda_r =1.5 \AA$ and a bunch length of 6.4 fs the maximal power increased from $\text{P}_{max}= 1.56$ TW for the Gaussian beam, to $\text{P}_{max}= 2.26$ TW for a parabolic beam and $\text{P}_{max}= 2.63$ TW for a uniform beam. An empirical argument related to the growth of the sideband instability has been considered to explain this discrepancy in maximal power output. It was found that for $\lambda_r =1.5 \AA$  in the transversely Gaussian beam case the energy deposited in the sidebands was greater by as much as 35 $\%$ compared to the parabolic case and 47 $\%$ compared to the uniform case. This effect is enhanced when one considers detuning the central beam wavelength off resonance and in the $\lambda_r =1.50078 \AA$ case the discrepancy between Gaussian and parabolic is $61 \%$ and it is $72 \%$ between Gaussian and uniform beams. 

Furthermore, a numerical study was made to recreate the small-signal high-gain curve for a tapered X-FEL and compare it to established theoretical results for fixed parameter undulators \citep{murphypellegrini}. It was found that redshifting the beam resonant wavelength by $\Delta \lambda / \lambda < 2\rho $ increased the maximum output power for all three transverse distributions. The maximum output power was obtained for a transversely flat beam at a value $\text{P}_{max}= 2.7$ TW at $\Delta \lambda / \lambda = 0.34 \rho$. 

This study has shown that that transverse pulse shaping may be chosen to improve the performance and increase the output power of a tapered free electron laser. More detailed simulations and complex taper profiles will be investigated in the future to determine the scope of these transverse effects on the output power of a tapered FEL. In light of the promising results found in this study we propose to investigate experimental realizations of transversely shaped electron beam distributions in order to measure the effects in the lab and compare the results with simulations. The use of suitable masks inside the beamline, or a number of more sophisticated methods, like the introduction of nonlinear elements in the electron transport line, could be considered to generate and tailor the beam profile transversely. 


\bibliographystyle{ieeetr}
\bibliography{finalreportreferences.bib}

\end{document}